\documentclass[aps,showpacs,preprintnumbers,amsmath,amssymb]{revtex4}
\UseRawInputEncoding 
\usepackage{dcolumn}
\usepackage[dvips]{epsfig}

\usepackage{float}
\usepackage{graphics,epsfig}
\usepackage{graphicx}
\usepackage{multirow}
\usepackage{dcolumn}
\usepackage{bm}

\begin{document}
\baselineskip=0.8 cm
\title{\bf Shadow of a disformal Kerr black hole in quadratic degenerate higher-order scalar-tensor theories}

\author{Fen Long$^{1}\footnote{fenlong163@163.com}$
Songbai Chen$^{1,3}$\footnote{Corresponding author: csb3752@hunnu.edu.cn},
Mingzhi Wang$^{2}$\footnote{wmz9085@126.com}
Jiliang Jing$^{1,3}$ \footnote{jljing@hunnu.edu.cn}}
\affiliation{ $ ^1$ Department of Physics, Key Laboratory of Low Dimensional Quantum Structures
and Quantum Control of Ministry of Education, Synergetic Innovation Center for Quantum Effects and Applications, Hunan
Normal University,  Changsha, Hunan 410081, People's Republic of China
\\
$ ^2$ School of Mathematics and Physics, Qingdao University of Science and Technology,
Qingdao, Shandong 266061, People¡¯s Republic of China
\\
$ ^3$Center for Gravitation and Cosmology, College of Physical Science and Technology, Yangzhou University, Yangzhou 225009, People's Republic of China}

\begin{abstract}
\baselineskip=0.6 cm
\begin{center}
{\bf Abstract}
\end{center}

We have studied the shadow of a disformal Kerr black hole with an extra deformation parameter, which belongs to non-stealth rotating solutions in quadratic degenerate higher-order scalar-tensor (DHOST) theory. Our result show that  the size of the shadow increases with the deformation parameter for the black hole with arbitrary spin parameter. However, the effect of the deformation parameter on the shadow shape depends heavily on the spin parameter of black hole and the sign of the deformation parameter.
The change of the shadow shape becomes more distinct for the black hole with the more quickly rotation and the more negative deformation parameter. Especially, for the near-extreme black hole with negative deformation parameter, there exist a ``pedicel"-like structure appeared in the shadow, which increases with the absolute value of deformation parameter. The eyebrow-like shadow and the self-similar fractal structures also appear in the shadow for the disformal Kerr black hole in DHOST theory. These features in the black hole shadow originating from the scalar field could help us to understand the non-stealth disformal Kerr black hole and quadratic DHOST theory.

\end{abstract}

 \maketitle
\newpage
\section{Introduction}
The image of supermassive black hole in M87* \cite{fbhs1,fbhs6} together with the gravitational waves detected from binary black hole merger \cite{P1,P2,P3,P4,P5}
open a new era of testing gravity in strong field regimes. Einstein's general
relativity is found to be consistent with the current observations. However, the possibility of modified
gravity is not completely excluded at present. Thus, it is still necessary to investigate further the dynamical properties and
observational effects associated with gravitational field in other alternative theories of gravity.

It is well known that scalar-tensor theories are a kind of alternative theories of gravity, where there exists a scalar degree of freedom besides the gravitational metric. At present, the most general family of scalar-tensor
theories is the so-called degenerate higher-order scalar-tensor (DHOST) theories \cite{dhost1,dhost2,dhost3,dhost4}, which contain the higher order derivative of scalar field and satisfy with a certain set of degeneracy conditions. It is a further extension of Horndeski theories \cite{late2}  and Beyond Horndeski  theories \cite{late3}.  With the degeneracy conditions, the Ostrogradsky ghost can be vanished in the DHOST theories, even if there exist the higher-order equations of motion. It means that the degeneracy of Lagrangian \cite{dhost1,dhost2}, rather than the order of equations of motion, is the crucial element characterized higher-order theories owning only a single scalar degree of freedom. In general, it is very difficult to obtain exact solutions for black holes in alternative theories of gravity because the field equations become more complicated. However, some new black hole solutions in the DHOST theories have emerged these last years \cite{c2633,c2634,c2635,dhost24,dhost26,dhost20,dhost27,dhost28}.  These solutions can be classified as the stealth solutions and the non-stealth solutions. For the stealth solutions, the metric of spacetime is the same as those in general relativity, but there exists an extra scalar field which does not emerge in the spacetime metric. However, for the non-stealth solutions, the parameters of scalar field appear in the metric, which lead to that the metric for the solution deviate from those in Einstein's theory of gravity.

Recently, a disformal rotating black hole solution in quadratic DHOST theories is obtained by using the disformal solution-generating method \cite{dhostv1,dhostv2}. Starting from a ``seed" known solution $\tilde{g}_{\mu\nu}$ in DHOST Ia
theory \cite{dhost3}, one can obtain a new solution  $g_{\mu\nu}$ in another specific DHOST Ia theory by performing a disformal transformation of the metric. It is not surprising because the disformal and conformal transformations can take us from some DHOST Ia theory to some other specific DHOST Ia theory \cite{dhost3}. This disformal rotating black hole solution owns three parameters: the mass, the spin and the disformal parameter which describes the deviation from the Kerr geometry \cite{dhostv1,dhostv2}. The scalar field associated with this solution is time-dependent with a constant kinetic density. The disformed Kerr solution is asymptotically flat, but no more Ricci flat. Especially, it has the same singularity as in the Kerr case despite the extra scalar field changes the metric of spacetime.
However, the presence of scalar field modifies the position and width of ergoregions, and then also affects the possibility of extracting rotation energy from black hole by the Penrose process. Therefore, comparing with the stealth solution, the change of spacetime structure originating from the scalar field for a rotating non-stealth black hole would bring us some new observational effects differed from usual Kerr black hole.

On the other hand, the investigations of black hole shadows have been boosted by the first image provided by the Event Horizon Telescope (EHT) Collaboration \cite{fbhs1,fbhs6}. It is shown that the information stored in shadows could be applied  to examine the theories of gravity. In general relativity, the shadow for a fast rotating black hole is
a ``D"-shaped silhouette due to dragging effect arising from black hole rotation \cite{kerr1,kerr2}. However, in alternative theories of gravity, the shadow for a fast rotating black hole could become more prolate or oblate due to the deviation from the Kerr black hole \cite{eht1,nkerr}.
Thus, with high precision observation in the future, it is possible to select out the theory describing correctly gravity in nature through detecting  black hole shadows. Moreover, black hole shadows have also been treated as a potential tool to study the possibility of constraining black hole parameters and extra dimension size \cite{extr1,extr2}, and to probe some fundamental physics issues including dark matter \cite{tomoch,dark1,dark2,dark3,dark4} and the equivalence principle \cite{epb}. The main purpose of this paper is to study the shadow of a rotating non-stealth black hole in quadratic DHOST theories and to see what new features exist in the shadow for this disformed black hole.

The paper is organized as follows: In Sec.II,  we introduce briefly the rotating non-stealth black hole solution in quadratic DHOST theories \cite{dhostv1,dhostv2}, and then study equation of motion for the photons in this spacetime. In Sec.III, we present numerically the shadow of the rotating disformed black hole and probe the effects of the deformation parameter arising from scalar field on the shadow. Finally, we end the paper with a summary.

\section{Equation of motion for the photons in the rotating non-stealth solution in quadratic DHOST theories}

Let us now to introduce briefly the disformal Kerr black hole with an extra deformation parameter, which is a non-stealth rotating solution in quadratic DHOST theory. The most general action for the quadratic DHOST theory  can be expressed as \cite{dhost1,dhost2,dhost3,dhost4}
\begin{eqnarray}
S=\int d^{4} x \sqrt{-g}\left(P(X, \phi)+Q(X, \phi) \square \phi+F(X, \phi) R+\sum_{i=1}^{5} A_{i}(X, \phi) L_{i}\right),\label{S}
\end{eqnarray}
with
\begin{eqnarray}
L_{1} &\equiv& \phi_{\mu \nu} \phi^{\mu \nu}, \quad L_{2} \equiv(\square \phi)^{2}, \quad L_{3} \equiv \phi^{\mu} \phi_{\mu \nu} \phi^{\nu} \square \phi, \nonumber\\
L_{4} &\equiv& \phi^{\mu} \phi_{\mu \nu} \phi^{\nu \rho} \phi_{\rho}, \quad L_{5} \equiv\left(\phi^{\mu} \phi_{\mu \nu} \phi^{\nu}\right)^{2},
\end{eqnarray}
where  $R$ is the usual Ricci scalar and $\phi$ is the scalar field with kinetic term  $X \equiv \phi_{\mu} \phi^{\mu}$, and the quantity $\phi_{\mu} \equiv \nabla \phi$. Here $P$, $Q$, $F$ and  $A_i$ are functions of $\phi$ and $X$. In order to ensure only an extra scalar degree of freedom in addition to the usual tensor modes of gravity in quadratic DHOST theory, the functions $F$, $A_i$ must satisfy the so-called degeneracy conditions, but $P$ and $Q$ are  totally free. The degeneracy conditions for quadratic DHOST Ia theory is given in Ref.\cite{dhost3}. From the previous discussion, one can obtain a new solution  from  a ``seed" known solution  in DHOST Ia theory  by performing a disformal transformation of the metric.
In general, the  disformal transformation of the metric  can be expressed as \cite{dhost3}
\begin{eqnarray}
g_{\mu \nu}=A(X, \phi) \tilde{g}_{\mu \nu}-B(X, \phi) \phi_{\mu} \phi_{\nu},\label{metric1}
\end{eqnarray}
where $g_{\mu \nu}$ is the ``disformed" metric and $\tilde{g}_{\mu \nu}$ is the original ``seed" one. $A(X,\phi)$ and $B(X,\phi)$ are  conformal and disformal factors, respectively.
In order to obtain a new solution,  the functions $A$ and $B$ must satisfy the conditions that the two metrics are not degenerate. Starting from Kerr metric and adopting the transformation with $A(X, \phi)=1$ and $B(X, \phi)=B_0$ ($B_0$ is a constant), one can obtain the disformal Kerr metric \cite{dhostv1,dhostv2}
\begin{eqnarray}
ds^{2}&=&-\frac{\Delta}{\rho}(d t-a\sin^{2}\theta d\psi)^{2}+\frac{\rho}{\Delta} dr^{2}+\rho d\theta^{2}+\frac{\sin ^{2}\theta}{\rho}(adt-(r^{2}+a^{2})d\psi)^{2} \nonumber\\
&&+\alpha(dt+\sqrt{2Mr(r^{2}+a^{2})}/\Delta dr)^{2}.
\label{metric}
\end{eqnarray}
Here the scalar field is only a function of the coordinates $t$ and $r$, i.e.,
\begin{eqnarray}
&&\phi(t, r)=-m t+S_{r}(r), \quad S_{r}= -\int  \frac{\sqrt{\mathcal{R}}}{\Delta} d r, \nonumber\\
&&\mathcal{R}=2 M m^{2} r(r^{2}+a^{2}),\quad \Delta=r^{2}+a^{2}-2 M r.\label{scalar}
\end{eqnarray}
The parameter $\alpha$ is related to the rest mass $m$ of the scalar field by $\alpha=-B_{0}m^2$, and $\rho=r^2+a^2\cos{\theta}^2$.
The choice of $A_0=1$ can avoid a global physically irrelevant constant conformal factor in the metric. As in refs.\cite{dhost24,dhostv1,dhostv2}, here the scalar hair is assumed to has a constant standard kinetic term
\begin{eqnarray}
\tilde{g}^{\mu\nu}\phi_{\mu}\phi_{\nu}=-m^2.\label{scalar20}
\end{eqnarray}
This equation can be actually looked as the Hamilton-Jacobi equation of a timelike test particle with the rest mass $m$ and Hamilton-Jacobi potential $S=\phi$. It implies that the above scalar hair $\phi$ has a form of the Hamilton-Jacobi potential $S$ associated to the geodesic equation. Thus, the general form the scalar hair can be expressed as
\begin{eqnarray}
\phi(t, r, \theta, \psi)=-E t+L_z\psi+S_{r}(r)+S_{\theta}(\theta), \label{scalar2}
\end{eqnarray}
Here $E$, $L_z$ are two constants of the geodesic motion of the corresponding timelike test particle,  which are identified  with  the  energy  and  the  azimuthal  angular  momentum  respectively. The quantities $S_{r}(r)$ and $S_{\theta}(\theta)$ obey
\begin{eqnarray}
S_{r}=-\int\frac{\sqrt{\mathcal{R}}}{\Delta} dr, \quad\quad\quad \quad
S_{\theta}(\theta)=\int\sqrt{\Theta} d\theta,\label{scalar21}
\end{eqnarray}
with
\begin{eqnarray}
\mathcal{R}&=&[E(r^2+a^2)-aL_z]^2-\Delta [\mathcal{Q}+(aE-L_z)^2+m^2r^2], \nonumber\\
\Theta&=&-\sin^2\theta \bigg(aE-\frac{L_z}{\sin^2\theta}\bigg)^2+[\mathcal{Q}+(aE-L_z)^2-m^2a^2\cos^2\theta].
\end{eqnarray}
The regularity at the poles on the axes requires $\partial S/\partial \theta\rightarrow 0$ as $\theta\rightarrow 0,\pi$, which means \cite{dhost24,dhostv1,dhostv2} that $L_z=0$ and $\mathcal{Q}+(aE-L_z)^2=m^2a^2$.
Moreover, in order that the scalar hair can be well defined from the event horizon up to asymptotic infinity, one must
take $E=m$ so that particles can marginally reach timelike infinity \cite{dhost24,dhostv1,dhostv2}, which implies $\mathcal{Q}=0$ and $\Theta=0$. Thus, the scalar field $\phi$ is taken to be independent of the angular variables $\theta$ and $\psi$  \cite{dhostv1,dhostv2}, and then has a form (\ref{scalar}). The negative sign in $S_r$ was chosen because the scalar field should be regular at the horizons of a Kerr black hole $\Delta=0$ \cite{dhostv1}. Obviously, the disformal Kerr metric (\ref{metric}) in quadratic DHOST theory owns three parameters:  the mass parameter $M$ and spin parameter $a$ together with a new deformation parameter $\alpha$ encoded precisely the deviations
from general relativity. The presence of $\alpha$ means that the disformal Kerr metric (\ref{metric}) is a non-stealth solution in quadratic DHOST theory because the scalar field modifies the metric of spacetime, which is difference from that in the stealth solution case. Like the usual Kerr metric, the disformal Kerr metric (\ref{metric}) has a coordinate singularity at $\Delta=0$ and an intrinsic ring singularity at $\rho=0$. However, it must be pointed out that although the disformal Kerr metric (\ref{metric}) is asymptotically flat, its asymptotical behavior is not entirely the same as that of the Kerr one due to  the existence of $drdt$ term. Moreover, the disformal Kerr metric (\ref{metric}) is no longer Ricci flat, i.e., $R_{\mu\nu}\neq0$. Especially, the presence of the $drdt$ term also leads to the lack of circularity  in the disformal Kerr spacetime \cite{dhostv1,dhostv2}. It is different from that in the usual Kerr spacetime in general relativity because the Kerr spacetime is circular, i.e., the spacetime can be foliated by 2-surfaces (called meridional surfaces) everywhere orthogonal to the Killing field $\xi=\partial_t$ and $\eta=\partial_{\psi}$ \cite{circle1,circle2,circle3}. The lack of circularity modifies the structure of the black hole horizons so that the horizons depend on the polar angle $\theta$ and cannot be given by $r=const$ in Boyer-Lindquist coordinates, and then the corresponding surface gravity is no longer a constant \cite{dhostv1,dhostv2}.

We are now in position to study the motion of photon in the disformal Kerr black hole spacetime (\ref{metric}).
The Hamiltonian of a photon moving along null geodesics in the curved spacetime can be expressed as
\begin{eqnarray}
 H(x,p)=\frac{1}{2}g^{\mu \nu}(x)p_{\mu}p_{\nu}=0.\label{hamiltonian}
\end{eqnarray}
It is obvious that the metric functions in the disformal Kerr spacetime (\ref{metric}) are independent of the coordinates $t$ and $\psi$, so there exist two conserved quantities, i.e., the photon's energy $E_0$ and its z-component of the angular momentum $L_{z0}$.  However, due to the existence of $drdt$ term, the forms of $E_0$ and $L_{z0}$ are modified as
\begin{eqnarray}
E_0=-p_{t}=-g_{tt}\dot{t}-g_{tr}\dot{r}-g_{t\psi}\dot{\psi}, \quad L_{z0}=p_{\phi}=g_{t\psi}\dot{t}+g_{\psi\psi}\dot{\psi}.\label{conserved quantities}
\end{eqnarray}
With these two conserved quantities, one can obtain the equations of null geodesics
\begin{eqnarray}
\dot{t}&=&\frac{g_{\psi\psi}E_0+g_{t\psi}L_{z0}+g_{tr}g_{\psi\psi}\dot{r}}{g_{t\psi}^2-g_{tt}g_{\psi\psi}},\label{u1}\\
\dot{\psi}&=&\frac{g_{t\psi}E_0+g_{tt}L_{z0}+g_{tr}g_{t\psi}\dot{r}}{g_{tt}g_{\psi\psi}-g_{t\psi}^2},\label{u4}\\
\ddot{r}&=&\frac{g_{t\psi}^2-g_{tt}g_{\psi\psi}}{g_{rr}g_{t\psi}^2-g_{tt}g_{rr}g_{\psi\psi}+g_{tr}^2g_{\psi\psi}}
\bigg\{\frac{1}{2}(g_{tt,r}\dot{t}^2-g_{rr,r}\dot{r}^2+g_{\theta\theta,r}\dot{\theta}^2+g_{\phi\phi,r}\dot{\phi}^2
+2g_{t\psi,r}\dot{t}\dot{\psi}-2g_{\theta\theta,\theta}\dot{r}\dot{\theta})\nonumber\\
&+&\frac{g_{tr}}{g_{t\psi}^2-g_{tt}g_{\psi\psi}}\bigg[g_{t\psi}(g_{t\psi,r}\dot{t}\dot{r}
+g_{t\psi,\theta}\dot{t}\dot{\theta}+g_{\psi\psi,r}\dot{r}\dot{\psi}+g_{\psi\psi,\theta}\dot{\theta}\dot{\psi})\nonumber\\
&-&g_{\psi\psi}(g_{tt,r}\dot{t}\dot{r}+g_{tt,\theta}\dot{t}\dot{\theta}+g_{tr,r}\dot{r}^2
+g_{t\psi,r}\dot{r}\dot{\psi}+g_{t\psi,\theta}\dot{\theta}\dot{\psi})\bigg]\bigg\},\label{uu2}\\
\ddot{\theta}&=&\frac{1}{2g_{\theta\theta}}(g_{tt,\theta}\dot{t}^2+g_{rr,\theta}\dot{r}^2-g_{\theta\theta,
\theta}\dot{\theta}^2+g_{\psi\psi,\theta}\dot{\psi}^2+2g_{t\psi,\theta}\dot{t}\dot{\psi}-2g_{\theta\theta,r}
\dot{r}\dot{\theta}).
\label{uu3}
\end{eqnarray}
From Eqs. (\ref{u1}) and (\ref{u4}), one can find that the quantities $\dot{t}$ and $\dot{\psi}$ also depend on $\dot{r}$ due to the presence of the metric function $g_{tr}$, which means that the motion of photon has some behaviors differed from the case of Kerr black hole. Thus, it is expected that the  shadow of a disformal Kerr black hole should possess some new properties which do not belong to Kerr one.

\section{Shadow of the disformal Kerr black hole in quadratic DHOST theory}

Let us now to study the shadow casted by the disformal Kerr black hole (\ref{metric}) in quadratic DHOST theory and probe the influence of the deformation parameter $\alpha$ on the shadow. In the disformal Kerr black hole spacetime (\ref{metric}), the null geodesic equation (\ref{u1})-(\ref{uu3}) can not be variable-separable. Thus, we have to resort to "backward ray-tracing" method \cite{sw,swo,astro,chaotic,binary,sha18,my,BI,swo7,swo8,swo9,swo10} to simulate numerically the shadow. With this method, the light rays are assumed to evolve from the observer backward in time and then one can obtain the position of each pixel in the final image by solving numerically the null geodesic equations. The image of shadow in observer's sky is composed of the pixels connected to the light rays falling down into black hole.
Since the spacetime of the disformal Kerr black hole (\ref{metric}) is
asymptotically flat, one can expand the local basis of observer
$\{e_{\hat{t}},e_{\hat{r}},e_{\hat{\theta}},e_{\hat{\psi}}\}$ as a form in the coordinate basis of black hole spacetime $\{ \partial_t,\partial_r,\partial_{\theta},\partial_{\psi} \}$
\begin{eqnarray}
\label{zbbh}
e_{\hat{\mu}}=e^{\nu}_{\hat{\mu}} \partial_{\nu},
\end{eqnarray}
where the matrix $e^{\nu}_{\hat{\mu}}$ meets $g_{\mu\nu}e^{\mu}_{\hat{\alpha}}e^{\nu}_{\hat{\beta}}
=\eta_{\hat{\alpha}\hat{\beta}}$, and $\eta_{\hat{\alpha}\hat{\beta}}$ is the usual Minkowski metric. For an asymptotically flat stationary spacetime (\ref{metric}), it is convenient to choose a decomposition \cite{circle3,sw,swo,astro,chaotic,binary,sha18,my,BI,swo7,swo8,swo9,swo10,swo11}
\begin{eqnarray}
\label{zbbh1}
e^{\nu}_{\hat{\mu}}=\left(\begin{array}{cccc}
\zeta&\varepsilon&0&\gamma\\
0&A^r&0&0\\
0&0&A^{\theta}&0\\
0&0&0&A^{\psi}
\end{array}\right),
\end{eqnarray}
where $\zeta$, $\varepsilon$, $\gamma$, $A^r$, $A^{\theta}$,and $A^{\phi}$ are real coefficients. Actually, the decomposition (\ref{zbbh1}) is associated with a reference frame with zero axial angular momentum in relation to spatial infinity.
According to the Minkowski normalization
\begin{eqnarray}
e_{\hat{\mu}}e^{\hat{\nu}}=\delta_{\hat{\mu}}^{\hat{\nu}},
\end{eqnarray}
one can obtain
\begin{eqnarray}
\label{xs}
&&A^r=\frac{1}{\sqrt{g_{rr}}},\;\;\;\;\;\;\;
A^{\theta}=\frac{1}{\sqrt{g_{\theta\theta}}},\;\;\;\;\;\;\;
A^{\psi}=\frac{1}{\sqrt{g_{\psi\psi}}},\;\;\;\;\;\;\;
\zeta=\sqrt{\frac{g_{rr}g_{\psi\psi}}{g_{t\psi}^{2}g_{rr}+(g_{tr}^2-g_{tt}g_{rr})g_{\psi\psi}}},
\nonumber\\
&&\varepsilon=-\frac{g_{tr}}{\sqrt{g_{rr}}}\sqrt{\frac{g_{\psi\psi}}{g_{t\psi}^{2}g_{rr}+(g_{tr}^2-g_{tt}g_{rr})g_{\psi\psi}}},\;\;\;\;\;\;
\gamma=-\frac{g_{t\psi}}{\sqrt{g_{\psi\psi}}}\sqrt{\frac{g_{rr}}{g_{t\psi}^{2}g_{rr}+(g_{tr}^2-g_{tt}g_{rr})g_{\psi\psi}}}.
\end{eqnarray}
From Eq.(\ref{zbbh}), we find that the locally measured four-momentum $p^{\hat{\mu}}$ of a photon can be expressed as
\begin{eqnarray}
\label{dl}
p^{\hat{t}}=-p_{\hat{t}}=-e^{\nu}_{\hat{t}} p_{\nu},\;\;\;\;\;\;\;\;\;
\;\;\;\;\;\;\;\;\;\;\;p^{\hat{i}}=p_{\hat{i}}=e^{\nu}_{\hat{i}} p_{\nu}.
\end{eqnarray}
Combining with Eq.(\ref{xs}), we can obtain the locally measured four-momentum $p^{\hat{\mu}}$ in the disformal Kerr black hole spacetime (\ref{metric})
\begin{eqnarray}
\label{kmbh}
p^{\hat{t}}&=&\zeta E_0-\gamma L_{z0}-\varepsilon p_r,\;\;\;\;\;\;\;\;\;\;\;\;\;\;\;\;\;\;\;\;
p^{\hat{r}}=\frac{1}{\sqrt{g_{rr}}}p_{r},\nonumber\\
p^{\hat{\theta}}&=&\frac{1}{\sqrt{g_{\theta\theta}}}p_{\theta},
\;\;\;\;\;\;\;\;\;\;\;\;\;\;\;\;\;\;\;\;\;\;
p^{\hat{\psi}}=\frac{1}{\sqrt{g_{\psi\psi}}}L_{z0},
\end{eqnarray}
Thus,  the celestial coordinates for pixel corresponding to light ray in the spacetime (\ref{metric}) can be expressed as
\begin{eqnarray}
\label{xd1}
x&=&-r_{obs}\frac{p^{\hat{\psi}}}{p^{\hat{r}}}
=-r_{obs}\sqrt{\frac{g_{rr}}{g_{\psi\psi}}}\frac{g_{t\psi}\dot{t}+g_{\psi\psi}\dot{\psi}}{g_{rr}\dot{r}+g_{tr}\dot{t}}, \nonumber\\
y&=&r_{obs}\frac{p^{\hat{\theta}}}{p^{\hat{r}}}=
r_{obs}\frac{\sqrt{g_{rr}g_{\theta\theta}}\dot{\theta}}{g_{tr}\dot{t}+g_{rr}\dot{r}},
\end{eqnarray}
where $r_{obs}, \theta_{obs}$ are the radial coordinate and polar angle of observer.
\begin{figure}
\includegraphics[width=4.0cm ]{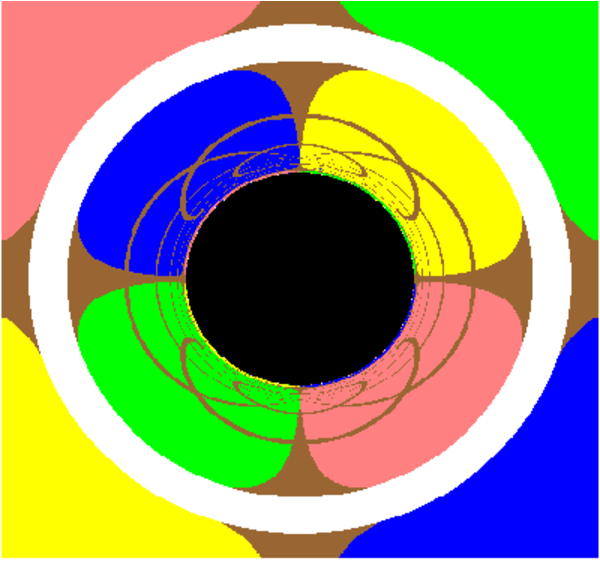}\includegraphics[width=4.0cm ]{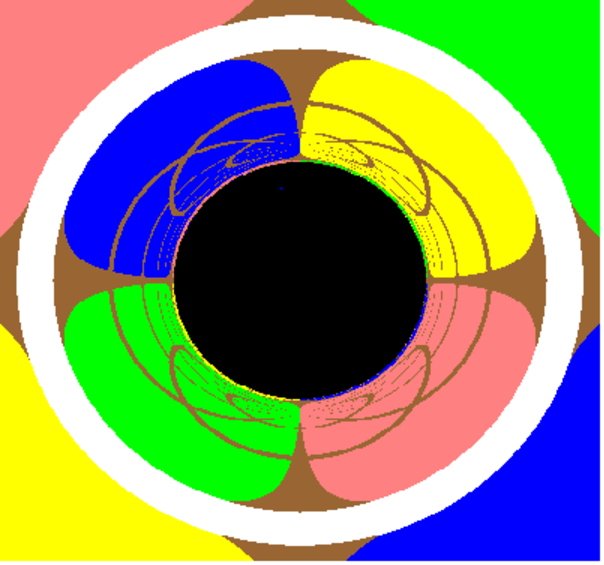}\includegraphics[width=4.0cm ]{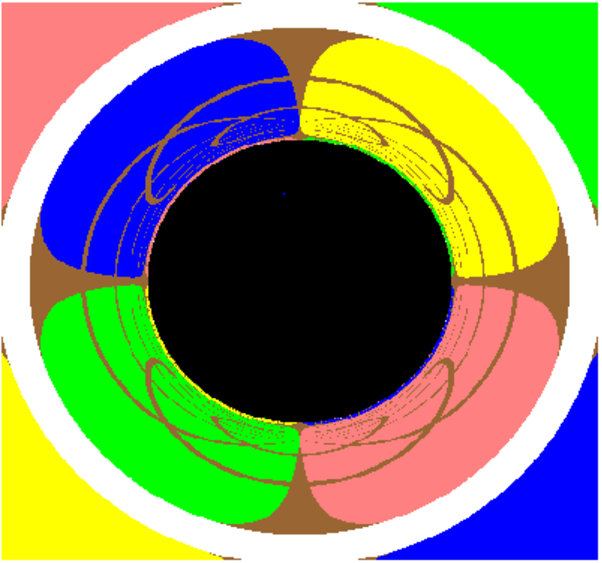}\includegraphics[width=4.0cm ]{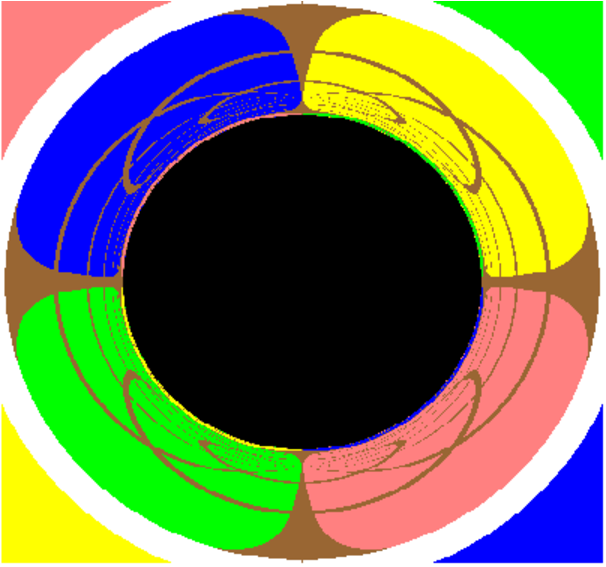}
\caption{The change of shadow with the deformation parameter $\alpha$ for the disformal Kerr black hole in the quadratic DHOST theory with fixed $a=0$. Here we set the mass parameter $M=1$, $r_{obs}=30M$ and $\theta_{obs}=\pi/2$. The figures from left to right correspond to $\alpha=-0.5$, $-0.3$, $0$, and $0.2$, respectively.}
\label{as0}
\end{figure}
\begin{figure}
\includegraphics[width=4.0cm ]{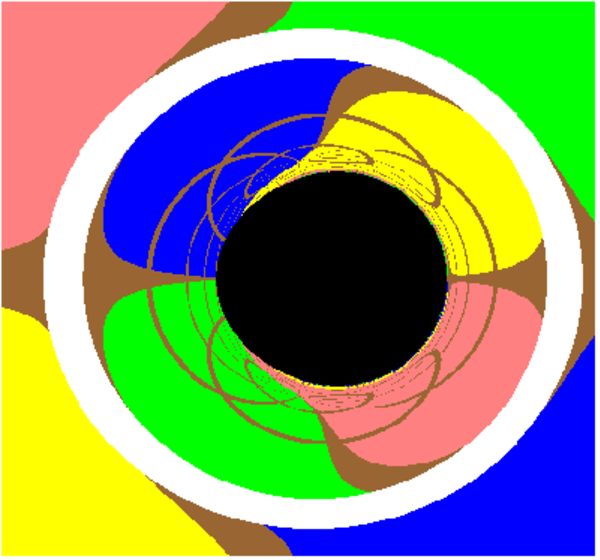}\includegraphics[width=4.0cm ]{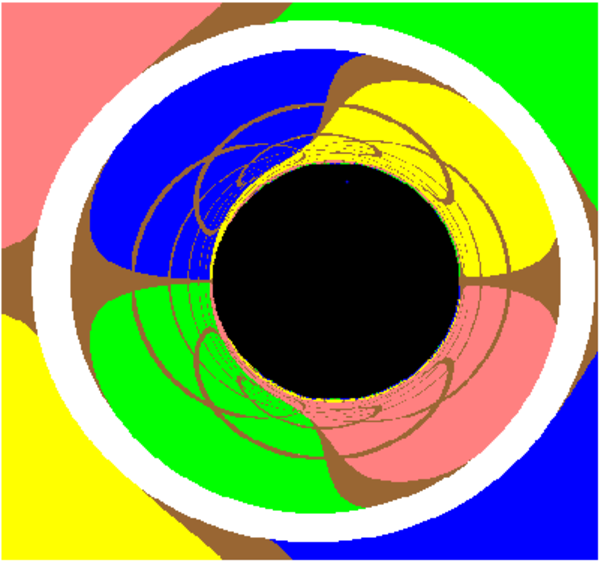}\includegraphics[width=4.0cm ]{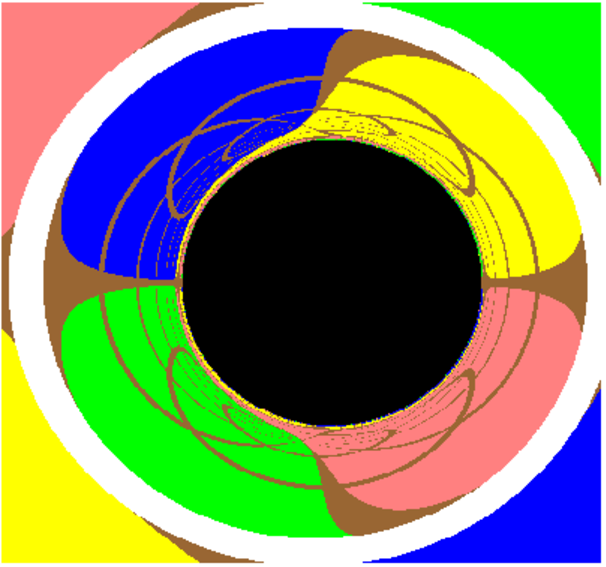}\includegraphics[width=4.0cm ]{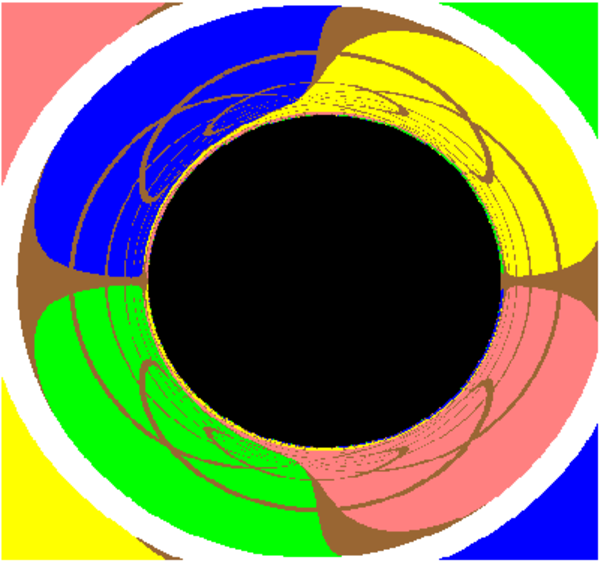}
\caption{The change of shadow with the deformation parameter $\alpha$ for the  disformal Kerr black hole in the quadratic DHOST theory with fixed $a=0.5$. Here we set the mass parameter $M=1$, $r_{obs}=30M$ and $\theta_{obs}=\pi/2$. The figures from left to right correspond to $\alpha=-0.5$, $-0.3$, $0$, and $0.2$, respectively.}
\label{as1}
\end{figure}
\begin{figure}
\includegraphics[width=4.0cm ]{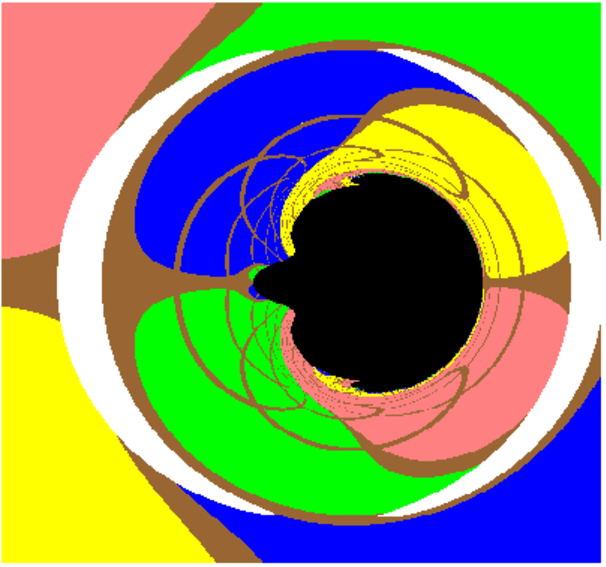}\includegraphics[width=4.0cm ]{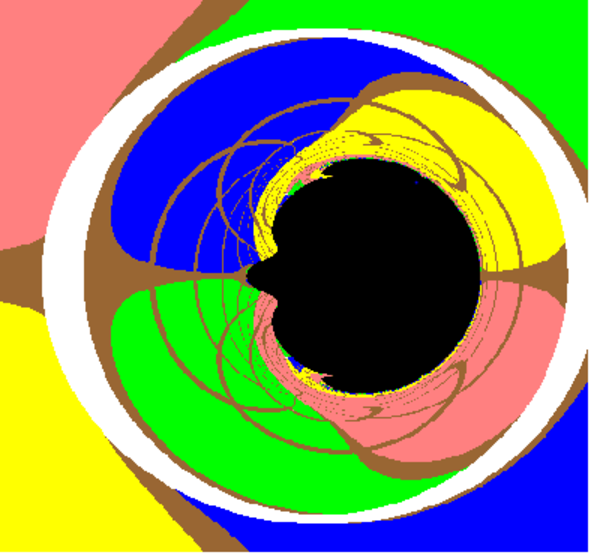}\includegraphics[width=4.0cm ]{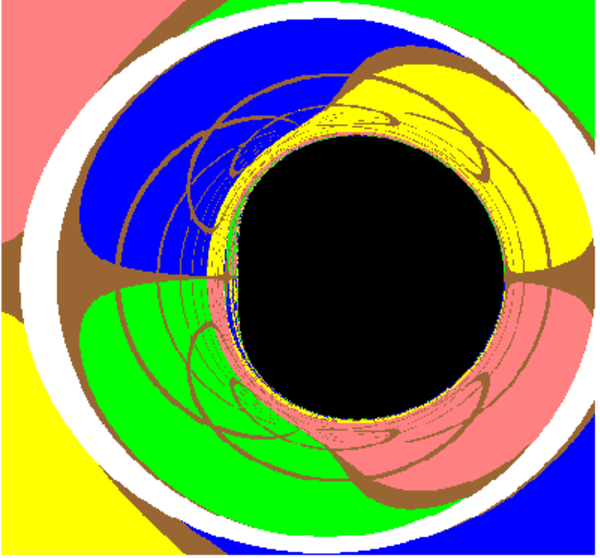}\includegraphics[width=4.0cm ]{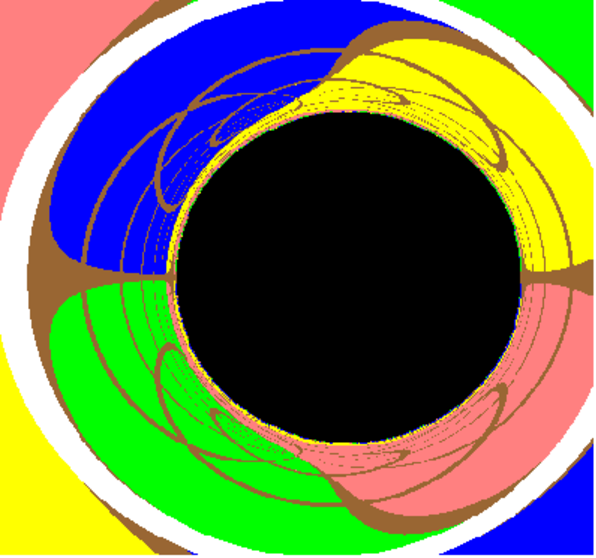}
\caption{The change of shadow with the deformation parameter $\alpha$ for the  disformal Kerr black hole in the quadratic DHOST theory with fixed $a=0.998$. Here we set the mass parameter $M=1$, $r_{obs}=30M$ and $\theta_{obs}=\pi/2$. The figures from left to right correspond to $\alpha=-0.5$, $-0.3$, $0$, and $0.2$, respectively.}
\label{as3}
\end{figure}

\begin{figure}
\quad\includegraphics[width=5.0cm ]{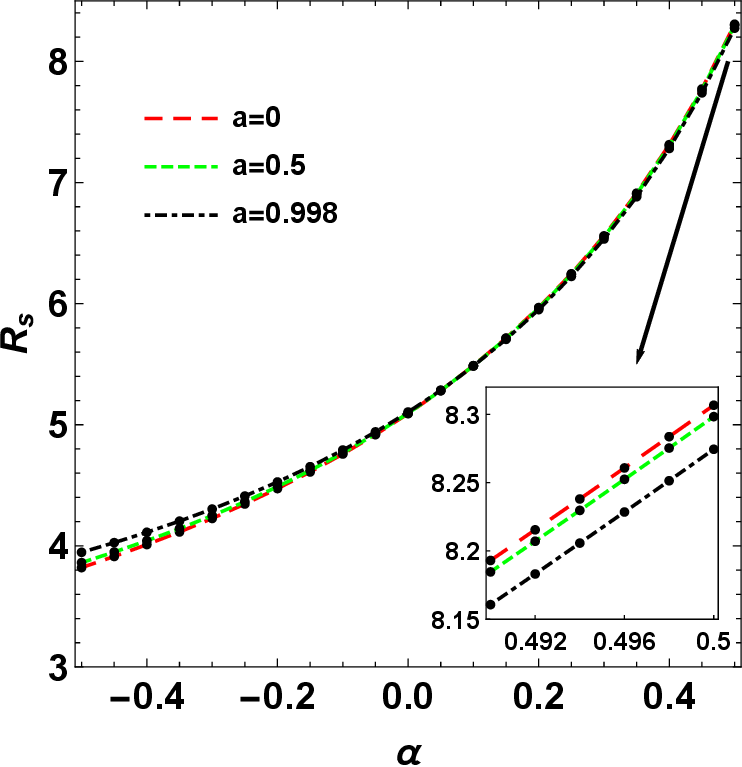}\quad\quad\includegraphics[width=5.4cm ]{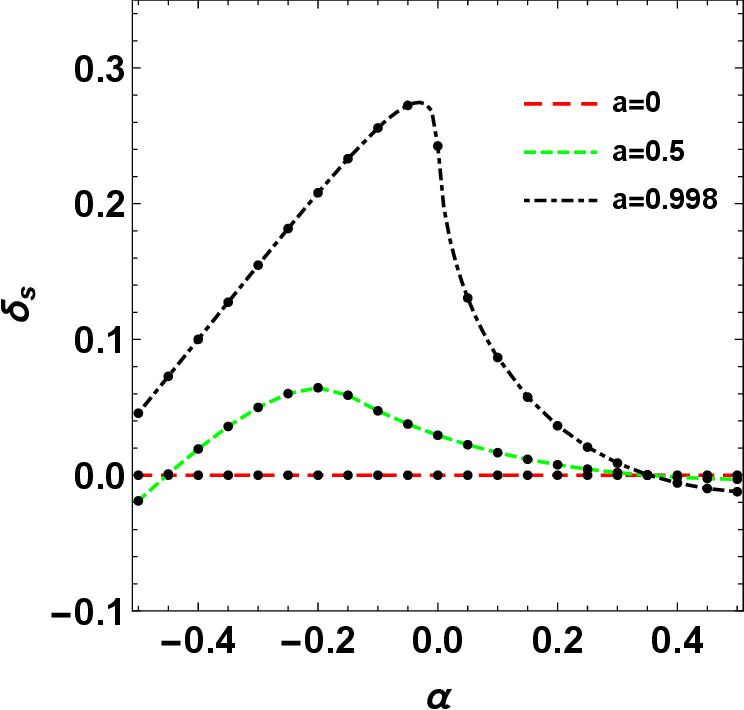}
\caption{The change of the shadow size $R_s$ and the distortion $\delta_s$ with the deformation parameter $\alpha$ for the  disformal Kerr black hole with different $a$. }
\label{as40}
\end{figure}

In Figs.\ref{as0}-\ref{as3},  we present the shadows of the disformal Kerr black hole (\ref{metric}) observed in equatorial plane for different spin parameters $a$ and deformation parameter $\alpha$. As in refs. \cite{sw,swo,astro,chaotic,binary,sha18,my,BI,swo7,swo8,swo9,swo10}, we
divide the celestial sphere into four quadrants painted in different colors (green, blue,
red, and yellow). The grid of longitude and latitude lines is marked with adjacent brown
lines separated by $10^\circ$. The distribution of these color regions and lines in figures reflects the
distortion of an image arising from the strong gravitational lensing of black hole. The black
part is used to denote black hole shadow. The white regions in figures are determined by lights from a reference spot lied in the line between black hole and observer, which provides a direct demonstration of Einstein ring.
As $a=0$, we find that the shadow shape is an perfect disk for different $\alpha$, which is similar to those of the usual non-rotating black holes. However, the size of the shadow increases with the deformation parameter $\alpha$. The dependence of the size of the shadow on the mass of the scalar field $m$ is also determined by the sign of the constant $B_0$. In other words,  the size of the shadow decrease with the mass $m$ of the scalar field as $B_0>0$, but increases as  $B_0<0$.
For the rotating black hole, from Fig.\ref{as1}-\ref{as3}, we find that effect of the deformation parameter $\alpha$ on the size of the shadow is similar to that in the non-rotating case, i.e., the size of the shadow is an increasing function of $\alpha$. However, the change of the shadow shape with $\alpha$ depends on the spin parameter $a$ of black hole and the sign of $\alpha$.
It is shown that the change of the shadow shape becomes more distinct for the black hole with the more quickly rotation and the more negative parameter $\alpha$. Especially, as $a=0.998$ and $\alpha<0$, we find that there exist a ``pedicel"-like structure appeared in the left of the shadow, which increases with the absolute value of $\alpha$.
In Fig.\ref{as40}, we also plot the change of the black hole shadow size parameter $R_s$ and the distortion parameter $\delta_s$ with the deformation parameter $\alpha$ for different spin parameter $a$. Here, the parameters $R_s$ and $\delta_s$ are defined as \cite{rs}
\begin{eqnarray}
R_s=\frac{(x_t-x_r)^2+y^2_t}{2(x_r-x_t)},\quad\quad\quad\quad\delta_s=2-\frac{x_r-x_l}{R_s},
\end{eqnarray}
where $(x_t, y_t)$ is the coordinate of the top point of the shadow. The coordinates $(x_r, 0)$ and $(x_l,0)$,  respectively, correspond to points located at the most right position and the most left position of shadow along the horizontal line $y=0$.
\begin{table}[ht]
\centering
\begin{tabular}{|c|c|c|c|c|c|c|c|c|c|c|c|c|}
\hline
\hline
\multicolumn{2}{|c|}{\multirow{2}{*}{$\delta_s$ }}&\multicolumn{11}{c|}{$\alpha$ }\\
\cline{3-13}\multicolumn{2}{|c|}{}&$-0.5$ &$-0.4$&$-0.3$&$-0.2$&$-0.1$ &$0.0$&$0.1$&$0.2$&$0.3$&$0.4$&$0.5$\\
\hline
\multirow{2}{*}{$a$ }& $0.5$ & $-0.0189$ & $0.0193$ & $0.0499$ & $0.0644$ & $0.0474$ & $0.0294$ & $0.0166$ & $0.0077$ & $0.0019$ & $-0.0015$ & $-0.0030$ \\
\cline{2-13}  & $0.998$ & $0.0457$  & $0.1001$& $0.1547$ & $0.2081$ & $0.2558$ & $0.2425$ & $0.0867$ & $0.0364$ & $0.0089$ & $-0.0058$ & $-0.0121$\\
\hline\hline
\end{tabular}
\caption{Numerical values of distortion parameters $\delta_{s}$ for the shadow casted by a disformal Kerr black hole in the quadratic DHOST theory. }
\label{table1}
\end{table}
From Fig.\ref{as40} and Tab.(\ref{table1}), we find that the shadow size $R_s$ increases with the increase of the deformation parameter $\alpha$ for fixed spin parameter $a$, which is consistent with the previous analysis. The change of $R_s$ with $a$ also depends on the disformal parameter $\alpha$. With the increase of $\alpha$, the shadow size $R_s$ change gradually from an increasing function of spin parameter $a$ to an decreasing function. For the distortion parameter $\delta_{s}$, one can find that
it first increases and then decreases with $\alpha$ for the fixed non-zero spin parameter $a$. Thus, there exists a peak in the curve $\delta_{s} (\alpha)$ and the position of peak moves right with the increase of $a$. From Fig.\ref{as40}, we confirm again that the change of the shadow shape with $\alpha$ becomes more distinct for the rapidly rotating  black hole.
\begin{figure}
\includegraphics[width=4.25cm ]{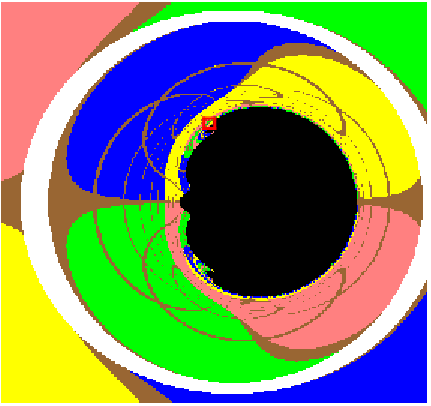}\includegraphics[width=4.0cm ]{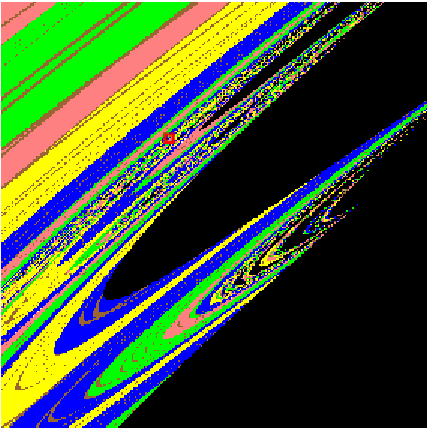}\includegraphics[width=4.0cm ]{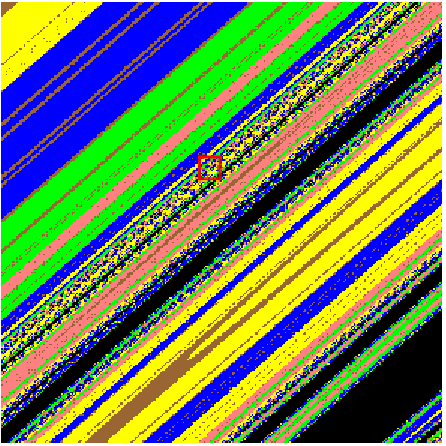}\includegraphics[width=4.0cm ]{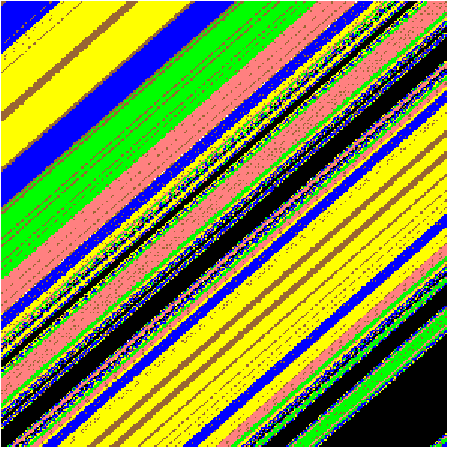}
\caption{The eyebrow shape shadow and the self-similar fractal structures in the shadow of a disformal Kerr black hole in the quadratic DHOST theory with fixed $a=0.998$ and $\alpha=-0.1$. Here we set the mass parameter $M=1$, $r_{obs}=30M$ and $\theta_{obs}=\pi/2$.}
\label{as5}
\end{figure}
Moreover, from Fig.\ref{as3}, we can find that as $\alpha<0$ there exist two larger eyebrow-like silhouette in the shadow, which distribute symmetrically on the two side of the horizontal line $y=0$. The region of appearing eyebrow-like shadow increases with the absolute value of $\alpha$. Actually, as shown in Fig.(\ref{as5}), many other smaller eyebrow-like structures can be detected in the shadow of the disformal Kerr black hole (\ref{metric}).  This means that the shadow of a disformal Kerr black hole in DHOST theory possesses a self-similar fractal structure caused by chaotic scattering  of photon, which is qualitatively different from that in usual Kerr black hole spacetime. These novel features  in the black hole shadow originating from the scalar field could help us to understand black hole and to examine whether the real gravity is described by quadratic DHOST theory in the future.

\section{summary}

We have studied the shadows of a disformal Kerr black hole with an extra deformation parameter $\alpha$, which belongs to  non-stealth rotating solutions in quadratic DHOST theory. Our result show that  the size of the shadow for fixed $a$ increases with the deformation parameter $\alpha$, which means that the size of the shadow decrease with the mass $m$ of the scalar field as $B_0>0$, but increases as  $B_0<0$. However, the effect of the parameter $\alpha$ on the shadow shape depends heavily on the spin parameter of black hole and the sign of $\alpha$. As $a=0$, the shadow shape is an perfect disk for different $\alpha$, which is similar to those of the usual non-rotating black holes. For the rotating case, the change of the shadow shape becomes more distinct for the black hole with the more quickly rotation and the more negative parameter $\alpha$. Especially, as $a=0.998$ and $\alpha<0$, we find that there exist a ``pedicel"-like structure appeared in the left of the shadow, which increases with the absolute value of $\alpha$.
The distortion parameters $\delta_{s}$ first increases and then decreases with the deformation parameter $\alpha$ for the fixed non-zero spin parameter $a$. The position of peak in the curve $\delta_{s} (\alpha)$ moves right with the increase of the spin parameter $a$. With the increase of the spin parameter $a$, we also note that the shadow size $R_s$ decreases for certain cases with positive $\alpha$, which is different from that in usual rotating black hole.
Moreover, we find that there exist eyebrow-like shadow and the self-similar fractal structures in the shadow for the disformal Kerr black hole in DHOST theory, which is qualitatively different from that in usual Kerr black hole spacetime.

\section{\bf Acknowledgments}

This work was  supported by the National Natural Science
Foundation of China under Grant No.11875026, 11875025 and 12035005, the Scientific Research Fund of Hunan Provincial Education Department Grant No. 17A124.

\end{document}